\documentstyle[prl,aps]{revtex}
\input BoxedEPS.tex
\SetOzTeXEPSFSpecial
\HideDisplacementBoxes

\begin{document}

\twocolumn[\hsize\textwidth\columnwidth\hsize
           \csname @twocolumnfalse\endcsname
\title{Double-exchange is not the cause of ferromagnetism
in doped manganites}

\author{G. M.~Zhao}
~\\
~\\
\address{Physik-Institut der Universit\"at Z\"urich,
CH-8057 Z\"urich, Switzerland}

\maketitle
\noindent
\begin{abstract}
The coexistence of ferromagnetism and metallic
conduction in doped manganites has long been explained by a double-exchange
model in which the ferromagnetic exchange arises from the carrier
hopping. We evaluate the zero-temperature spin stiffness $D(0)$ and
the Curie temperature $T_{C}$ on
the basis of the double-exchange model using the measured values
of the bare bandwidth $W$ and the
Hund's rule coupling $J_{H}$. The calculated $D(0)$ and $T_{C}$
values are too small compared with the observed ones even in the
absence of interactions.
A realistic onsite interorbital Coulomb
repulsion can reduce $D(0)$ substantially in the case of a 2-orbital model.
Furthermore, experiment shows that $D(0)$ is simply proportional to $x$ in
La$_{1-x}$Sr$_{x}$MnO$_{3}$ system, independent of whether the ground
state is a ferromagnetic insulator or metal. These results strongly
suggest that the ferromagnetism in
manganites does not originate from the double-exchange interaction.
On the other hand, an alternative model based on the $d-p$ exchange
can semi-quantitatively explain the ferromagnetism of doped manganites at
low temperatures.
\end{abstract}
\vspace{1cm}
]

\narrowtext

The discovery of ``colossal" magnetoresistance in thin films of
the manganite perovskites Re$_{1-x}$D$_{x}$MnO$_{3}$ (Re = a
rare-earth element, and D = a divalent
element) \cite{Von,Jin} has stimulated extensive studies of magnetic,
structural and
transport properties of these
materials \cite{Art}. The coexistence of ferromagnetism and metallic
conduction has long been explained by the
double-exchange (DE) model \cite{Zener,Anderson}, where the effective hopping
for the manganese 3$d$ conduction electrons varies with the angle between
the manganese core electrons due to a strong Hund's coupling. However,
Millis {\em et al.} \cite{Millis1} proposed that, in addition to the
double-exchange, a strong electron-phonon interaction arising
from a strong Jahn-Teller effect should be involved to explain the
basic physics of manganites. In this modified model, the primary cause of
the ferromagnetism of doped manganites is still the double-exchange
interaction.

In the DE model, it is implicitly assumed that
doped carriers are Mn $e_{g}$ electrons. This assumption
is not justified by both electron-energy-loss \cite{Ju} and photoemission
spectroscopies
\cite{Saitoh}, which have shown that
the ferromagnetic manganites ($x$$<$0.4) are doped charge-transfer insulators
with carriers mainly residing on the oxygen orbitals. Now a question
arises: Does the ferromagnetism of
doped manganites really originate from the DE interaction?  If not,
what causes the ferromagnetism in these compounds? One way to address
this fundamental issue is to make a quantitative comparison between the
predicted properties of the DE model and experiment.

There are two important parameters in the DE model, namely,
the bare bandwidth $W$ of the $e_{g}$ bands, and the
Hund's rule coupling $J_{H}$ between $e_{g}$ and $t_{2g}$ electrons.
These parameters are related to an optical
transition between the exchange splitted $e_{g}$ bands
\cite{Furukawa,Chat}, and thus can be determined from optical data. With
these unbiased parameters, one can calculate the zero-temperature spin
stiffness
$D(0)$ and $T_{C}$. Here $D(0)$ is defined as $\omega_{q} = D(0)q^{2}$
with $\omega_{q}$ being the magnon
frequency.  When one introduces interactions such as electron-phonon
and electron-electron interactions, the magnitudes of both $D(0)$ and
$T_{C}$ are generally reduced.

Here we use the measured values of the bare bandwidth $W$ and the
Hund's rule coupling $J_{H}$ to calculate $D(0)$ and $T_{C}$ on the basis
of the double-exchange model. The calculated $D(0)$ and $T_{C}$
values are too small compared with the observed ones.
A realistic onsite interorbital Coulomb
repulsion can reduce $D(0)$ substantially in the case of a 2-orbital model.
Moreover, experimental data show that $D(0)$ is simply proportional to $x$ in
La$_{1-x}$Sr$_{x}$MnO$_{3}$ system, independent of whether the ground
state is a ferromagnetic insulator or metal. These results provide
strong evidence that the ferromagnetism in manganites is not
caused by the DE interaction. On the other hand, an alternative model
based on the $d-p$ exchange can well explain the
ferromagnetism of doped manganites.

Now we start with a Kondo-lattice type Hamiltonian \cite{Qui}, which
leads to Zener's DE model when $J_{H}$ $\rightarrow$ $\infty$,
\begin{eqnarray}
H = -\frac{1}{2}\sum_{<ij>ab\alpha} t_{ij}^{ab}
(d_{ia\alpha}^{\dagger}d_{jb\alpha} +
h.c.) \nonumber \\
-J_{H}\sum_{ia\alpha\beta}\vec{S}_{i}\cdot d_{ia\alpha}^{\dagger}
\vec{\sigma}_{\alpha\beta}d_{ia\beta} + H_{INT}.
\end{eqnarray}
Here $d_{ia\alpha}^{\dagger}$ creates an electron in $e_{g}$ orbital $a$
with spin $\alpha$,  $t_{ij}^{ab}$ is the direction-dependent
amplitude for an electron to hop from orbital $a$ to orbital $b$ on
a neigboring site, and $H_{INT}$ represents the other interactions.
The calculated band structure is well fit by a
$t_{ij}^{ab}$, which involves only nearest-neighbor hopping that is
only nonzero for one particular linear combination of orbitals, i.e.,
$t_{ij}^{ab}$ $\propto$ $t$, where $t$ is a characteristic hopping
amplitude that is related to the bare bandwidth $W$ by $W$ = 4$t$.
Here we still call Eq.~1 as
2-orbital DE model rather than 2-orbital
Kondo-lattice model for convienence.
The quantum and thermal average of the hopping term in Eq.~1,
defines a quantity $K$:
\begin{equation}
K = (1/6N_{site})
\sum_{<ij>ab\alpha}t_{ij}^{ab}\langle
d_{ia\alpha}^{\dagger}d_{jb\alpha} + h.c.\rangle,
\end{equation}
The quantity $K$ is related to the optical
spectral weight by a familar sum rule,
\begin{equation}
K = \frac{2a_{\circ}}{\pi e^{2}}\int_{0}^{\infty}d\omega \sigma_{1}(\omega),
\end{equation}
where $\sigma_{1}$ is the real-part optical conductivity contributed
only from the $e_{g}$ electrons, and $a_{\circ}$ is the lattice constant. The
quantity $K$ generally consists of the Drude part $K_{D}$, and
incoherent part $K_{I}$ which, in general, involves interband and
intraband optical transitions.
The Drude part $K_{D}$ can be related to the plasma frequency
$\Omega_{p}$ as
\begin{equation}
K_{D} = \frac{a_{\circ}}{4\pi e^{2}}(\hbar\Omega_{p})^{2}.
\end{equation}

On the basis of Eq.~1, Quijada {\em et al.} \cite{Qui}
showed that, to the order of $1/J_{H}$, the spin
stiffness $D(0)$ is,
\begin{equation}
D(0) =
\frac{Ka_{\circ}^{2}}{4S^{*}}[1 - \frac{\eta t^{2}}{J_{H}SK}],
\end{equation}
where $S$ = 3/2, $S^{*}$ = $S$ + $(1-x)/2$, and $\eta$ = 1.04 when
$H_{INT}$ = 0. The presence of interactions may change the value of $\eta$.
A similar result was obtained by Furukawa \cite{Furukawa} for an
1-orbital DE model
using the dynamical mean field method, but the value of $\eta$ is
doping dependent and less than 1.

From the above equations, one can calculate $K$ and $D(0)$ using realistic
values of the bare bandwidth $W$ and the
Hund's rule coupling $J_{H}$. Both the local density approximation
(LDA) \cite{Pick2} and ``constrained'' LDA \cite{Satpathy} calculations show
that $J_{H} \simeq$ 1.5 eV. The calculated $J_{H}$ value is very
close to the atomic values for 3d atoms. This is reasonable because $J_{H}$
is not
screened when the ion is put in a solid. The bare bandwidth $W$
cannot be calculated reliably for 3d-metal based compounds due to
a strong correlation effect. Fortunately, the values of both $W$ and $J_{H}$
can be determined from an optical
transition between the exchange splitted $e_{g}$ bands \cite{Furukawa,Chat}.
The peak position of this optical transition is about 2$J_{H}$, and
the width of the peak contains information about the bare
bandwidth \cite{Furukawa,Chat}. From the optical data of
Ref.~\cite{Qui,Machida}, one finds $J_{H} \simeq$ 1.6 eV and $W$ = 1.6-1.8
eV by comparing the data with the theoretical
predictions \cite{Furukawa,Chat}.
The value of $J_{H}$ obtained from the
optical data is in
excellent agreement with the calculated one. This implies that the feature
appeared at about 3 eV in the optical data indeed arises from
the optical transition between the exchange splitted $e_{g}$ bands.

The quantity $K^{\circ}$ for noninteracting 2-orbital model can be
evaluated when the bare bandwidth $W$ is known. Takahashi and Shiba
\cite{Takahashi} have calculated the optical conductivity using a tight
binding (TB) approximation of the band structure. From their
calculated result for the interband optical conductivity, we evaluate that
$K_{I}^{\circ}$ = 0.088$t$ for $x$ $\sim$ 0.3. One should also note
that the magnitude of $t$ defined in Ref.~\cite{Takahashi} is 1.5 times
smaller than the $t$ defined here. Since $K_{D}^{\circ}$ = 1.2$K_{I}^{\circ}$
\cite{Takahashi}, then $K_{D}^{\circ}$ = 0.106$t$ and $K^{\circ}$ =
0.194$t$. The LDA calculation
for a cubic and undistorted structure shows that \cite{Pick1}
$W$ = 3 eV and $\hbar\Omega_{p}^{\circ}$ = 1.9 eV. Using Eq.~4 and
$\hbar\Omega_{p}^{\circ}$ = 1.9 eV, one yields $K_{D}^{\circ}$ = 78.6 meV.
Since $t$ = $W$/4 = 0.75 eV, one readily finds that $K_{D}^{\circ}$ =
0.105$t$, in remarkably good agreement with that ($K_{D}^{\circ}$ =
0.106$t$) estimated from the
TB approximation. This justifies the relation  $K^{\circ}$ =
0.194$t$ obtained from the TB approximation. It is interesting to
compare the present result with those reported in Ref.~\cite{Chat} and
\cite{Qui}. In Ref.~\cite{Chat}, it is found
that $K^{\circ}$ = 0.34$t$ for $x$ = 0.3 using the dynamic mean field
method. In Ref.~\cite{Qui}, Quijada {\em et al.,} claimed
$K^{\circ}$ = 0.46$t$ for $x$ = 0.3, which might be true if
the two $e_{g}$ bands have the same dispersion. Therefore the quantity
$K^{\circ}$ is significantly
overestimated in Ref.~\cite{Chat} and \cite{Qui}.

When the Hund's coupling $J_{H}$ is turned on, the quantity $K$ is
reduced compared with $K^{\circ}$. For $J_{H}$ = $\infty$ and $x$ =
0.3, $K$ = 0.77$K^{\circ}$ \cite{Chat,Kubo}. It was also shown that
\cite{Chat} the reduction factor (0.77) is basically the same for
$J_{H}$ $\geq$ $t$. Therefore, we have $K$ = 0.147$t$ for $x$ = 0.3.
Using $t$ = 0.4 eV, $J_{H}$ = 1.6 eV, and $K$ = 0.147$t$, we yield
$D(0)$ = $-$25 mV \AA$^{2}$ from Eq.~5. The negative value of $D(0)$
implies that the ferromagnetism is not sustainable with these
unbiased parameters. It might be possible to have a small
positive $D(0)$ if one includes higher order terms in Eq.~5.
Nevertheless, the theoretical $D(0)$ value is
too small compared with the measured ones (160-190 mV
\AA$^{2}$) \cite{Martin,Baca,Lynn}.

Now we turn to the calculation of $T_{C}$ for $x$ = 0.3. For
an 1-orbital DE model with $J_{H}$ = $\infty$, the dynamic
mean field (DMF) calculation
shows that $T_{C}^{MF}$ = $0.078t/k_{B}$ \cite{Furukawa}, while
Monte Carlo simulations yield $T_{C}$ =
0.04$t/k_{B}$ \cite{Cald}, or $T_{C}$ =
0.03$t/k_{B}$ \cite{Yun}. This implies that the DMF calculation overestimates
$T_{C}$ by a factor of about 2 due to the neglection of fluctuations. We
would like to mention that the magnitude of $t$ defined in
Ref.~\cite{Cald} and \cite{Yun} is 3 times smaller than the $t$
defined here. For realistic parameters $t$ = 0.4
eV, $J_{H}$ = 4$t$ = 1.6 eV, the DMF calculation obtains
$T_{C}^{MF}$ = $0.038t/k_{B}$ = 180 K \cite{Furukawa}. Considering the fact
that the DMF
method can overestimate $T_{C}$ by a factor of 2, one has $T_{C}$ $\sim$
100 K. For the 2-orbital DE
model, the DMF calculation shows \cite{Chat} $T_{C}^{MF}$ = 0.07$t/k_{B}$ = 324 
K. Since the quantity $K$ calculated in Ref.~\cite{Chat} is overestimated by
a
factor of about 2 (as discussed above),  and the DMF method itself can
overestimate $T_{C}$ by about 2 times, the real $T_{C}$ should be about 100
K, which is comparable with
the value for the 1-orbital model. This is reasonable because the
bare values of $K^{\circ}$ for both 1- and 2-orbital models happen to
be very similar in the case of $x$ = 0.3. Therefore, both
the 1- and 2-orbital DE models cannot explain the
observed $T_{C}$ values with the unbaised parameters.

The above calculations have not taken into account any other
interactions such as electron-phonon interaction and
electron-electron correlation. The electron-phonon interaction can
substantially reduce the $K$ and thus $D(0)$ if the coupling constant
$\lambda$ = $E_{p}/2t$ is greatly larger than 1 \cite{Firsov,Capone}.
In reality, the polaron binding energy $E_{p}$ in
manganites is estimated to be about 1 eV \cite{Alexcond}. So
$\lambda$ $\simeq$ 1, which suggests that the electron-phonon
coupling does not lead to a sizable decrease in $D(0)$. On the other
hand, the electron-electron correlation in the 2-orbital model can
lead to a large reduction in $K$, as demonstrated by Horsch and
coworkers \cite{Horsch}. The parameter $U$ = $U'$ - $J_{ab}$ (where
$U'$ is the onsite interorbital Coulomb repulsion, and $J_{ab}$ is the
interorbital Hund's coupling) has a strong influence on the value of
$K$ \cite{Horsch}.
With realistic values of $U$ = 3 eV \cite{Horsch,Ishihara} and $t$ =
0.4 eV, $K$ can be reduced by a factor of about 2 compared with the bare
one \cite{Horsch}. Although their results are for 2-dimensional finite
clusters,
we would expect a
similar result for a real 3-dimensional system.
\begin{figure}[htb]
    \ForceWidth{7cm}
	\centerline{\BoxedEPSF{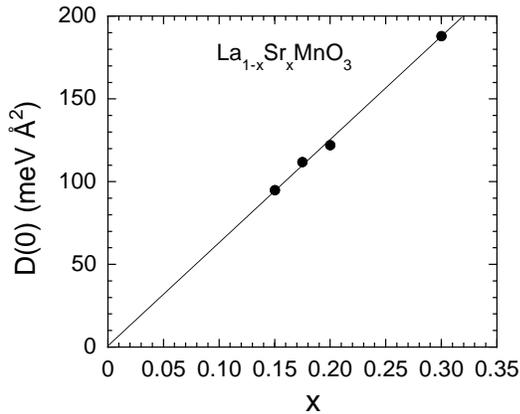}}
	\vspace{0.3cm}
	\caption[~]{Zero-temperature spin stiffness $D(0)$ vs $x$ in
	La$_{1-x}$Sr$_{x}$MnO$_{3}$ system. The $D(0)$ value for $x$ = 0.15
	is taken from Ref.~\cite{Doloc}, and the other data from
	Ref.~\cite{Hirota}. It is evident that $D(0)$ is simply proportional
	to $x$, independent of whether the ground state is a ferromagnetic
	insulator (e.g., $x$ = 0.15), or a ferromagnetic metal (e.g., $x$ =
	0.175, 0.20, 0.30). }
	\protect\label{Fig.1}
\end{figure}

In addition, it is striking that the observed spin stiffness varies only with
$x$, namely, $D(0) \propto x$, as seen clearly from Fig.~1.
The doping dependence of $D(0)$ shown in Fig.~1 could be
qualitatively explained by the 1-orbital DE model \cite{Furukawa,Varma}
where $D(0) \propto K \propto x(1-x)$ for $J_{H}$ = $\infty$ \cite{Varma},
as plotted in
Fig.~2. Nevertheless, the doping dependence of
$D(0)$ is very different from that prediced from the 2-orbital model.
By analogy to
the 1-orbital model, one can easily
show that $D(0) \propto K \propto
(1-x)(1+x)$ for the 2-orbital model, as demonstrated in Fig.~2. Here
we have implicitly assumed that the two $e_{g}$ bands have the same
dispersion.
The solid circles in Fig.~2
are the results evaluated using the dynamical mean field method with
$J_{H}$ = $\infty$ \cite{Chat}. It is
evident that the analytical
expression is in excellent agreement with the numerical result.
Comparing the results shown in Fig.~1 and Fig.~2, one clearly sees
that the 2-orbital model cannot explain the
observed doping dependence of $D(0)$.
\vspace{1cm}
\begin{figure}[htb]
    \ForceWidth{7cm}
	\centerline{\BoxedEPSF{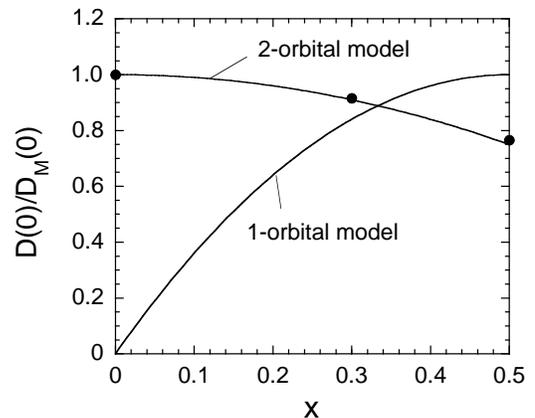}}
	\vspace{1cm}
	\caption[~]{The doping dependence of $D(0)$ predicted by the
	1-orbital and 2-orbital DE models with $J_{H}$ $\rightarrow$
$\infty$.  The
	solid lines represent $D(0)$ $\propto$ $x(1-x)$ for the 1-orbital
	model, and $D(0)$ $\propto$ $(1-x)(1+x)$ for the 2-orbital model.
	The solid circles are the values numerically calculated from
	the 2-orbital model \cite{Chat}.}
	\protect\label{Fig.2}
\end{figure}

The question is why the DE model cannot explain the
ferromagnetism in doped manganites. As mentioned above, the DE model implicitly 
assumes that doped carriers are Mn $e_{g}$ electrons, which is not 
the
case according to the electron-energy-loss and photoemission
spectra \cite{Ju,Saitoh}. Furthermore, the ``constrained'' LDA calculation
\cite{Satpathy} shows a
large onsite Coulomb repulsion of about 8-10 eV, in agreement with the
photoemission data \cite{Saitoh}. The simple LDA calculation which
ignores the strong correlation effect shows a bare plasma
frequency of about 1.3 eV for $x$ = 0.33 with a distorted structure
determined by neutron scattering \cite{Pick}. The bare plasma
frequency calculated is much smaller than
the one observed in Nd$_{0.7}$Sr$_{0.3}$MnO$_{3}$ (3.3 eV)
\cite{Zhaopreprint1}.
The large bare plasma
frequency observed in this material is consistent with the fact that
doped holes reside mainly on the oxygen orbitals with a large
bandwidth. The bare plasma frequency of about 3.3 eV for
single conduction band (oxygen band) implies a bare $K^{\circ}$ of
about 0.24 eV,
which gives an upper limit for the $K$ in the presence of interactions.
The electron-phonon interaction
with a coupling constant $\lambda$ $\sim$ 1 will reduce the $K$
slightly \cite{Firsov,Capone}, but can
significantly decrease the Drude
weight. Optical data indeed show that the
$K$ for Nd$_{0.7}$Sr$_{0.3}$MnO$_{3}$ is about 0.2 eV \cite{Qui}, while
the effective plasma frequency is about 0.57 eV \cite{Lee}. The Drude
weight is reduced by a factor of 33, implying small polaronic carriers
in the low-temperature ferromagnetic state.

\begin{figure}[h]
    \ForceWidth{7cm}
	\centerline{\BoxedEPSF{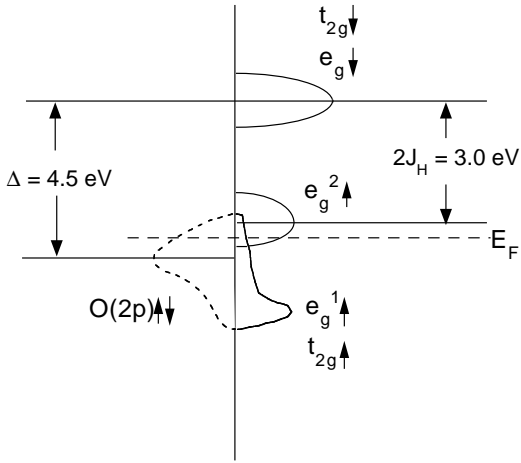}}
	\vspace{0.3cm}
	\caption[~]{The schematic band structure for doped manganites
	($x$$<$0.4) constructed from the LDA + U calculation  \cite{Satpathy}.
	The energy scales are consistent with the optical data
	\cite{Qui,Lee,Jung}. The density of the oxygen holes is equal to $x$
	per cell while the densities of the Mn$^{2+}$ and Mn$^{4+}$ ions
are the
	same due to the charge disproportion (2Mn$^{3+}$ $\rightarrow$
Mn$^{2+}$ +
Mn$^{4+}$) \cite{Hundley}.}
	\protect\label{Fig.3}
\end{figure}

In Fig.~3, we plot a schematic band
structure for doped manganites ($x$$<$ 0.4), which is extracted from
the LDA + U calculation. Here we have assumed that the local
Jahn-Teller distortions still survive upon doping, but the average
magnitude of the distortions decreases, in agreement with the experiments
\cite{Louca,Lanzara}.  The doping with a divalent element
shifts down the $e_{g}^{2}$ band due to the decrease of the
Jahn-Teller distortions. The density of the oxygen holes is equal to $x$
per cell, while the electron carrier density in the majority $e_{g}^{2}$
band (corresponding to the density of the Mn$^{2+}$ ions) is the same as
the hole
carrier density in the majority $e_{g}^{1}$ band (corresponding to the
density of the
Mn$^{4+}$ ions). In other words, the doping does not change the
average valence of the Mn ions, but lead to
the charge disproportion (2Mn$^{3+}$ $\rightarrow$ Mn$^{2+}$ +
Mn$^{4+}$). This is because the quench of the static Jahn-Teller
distortions by doping makes the Mn$^{2+}$-Mn$^{4+}$ pairs more stable
than the Mn$^{3+}$-Mn$^{3+}$ pairs ~\cite{Hundley}. The current band structure
is consistent with the optical transitions at the photon
energies of about 1.5 eV, 3.0 eV and 4.5 eV \cite{Qui,Lee,Jung}.
The optical spectral weight for the 4.5 eV
transition should be much larger than for the 1.5 eV transition, as
observed \cite{Jung}. This is because the unoccupied state density for the
former optical transition
(i.e., 3 minority $t_{2g}$ and 2 minority
$e_{g}$ states per cell) is at
least 5 times larger than that for the latter
one (i.e., less than 1 majority $e_{g}$ states
per cell). The optical transition at about 3 eV is related to the
transition between the exchange splitted $e_{g}$ bands.

What is an alternative model for the ferromagnetism in doped
manganites? If we consider an oxygen hole (spin 1/2) sitting in
between two Mn ions, an exchange interaction between the oxygen and Mn
spins ($d-p$ exchange) will lead to a ferromagetic interaction between
Mn spins \cite{Khomskii}. In this case, the ferromagnetic exchange
energy between two Mn spins is $J$ = $C(t_{pd}^{4}/\Delta_{ct}^{3})$,
where $C$ is a numerical factor of order of 1, $t_{pd}$ is a hybridization
matrix element between the $d$ and $p$ orbitals, and $\Delta_{ct}$ is a charge
transfer gap. The long-wave spin stiffness for a
cubic material is \cite{Perring} $D(0)$ =
$(S^{*}/3)\sum_{j}J_{ij}|\vec{R}_{i} - \vec{R}_{j}|^{2}$,
where $J_{ij}$ is the exchange energy between pairs of
spins at sites $\vec{R}_{i}$ and $\vec{R}_{j}$, and the sum is over
six neighboring sites.  Clearly, $J_{ij}$ =
$J$ if there is an oxygen hole in between two Mn ions, while $J_{ij}$ = 0
if not.
Since the oxygen hole density is equal to $x$ per cell,  then $D(0)$  =
$(2/3)xJS^{*}a_{\circ}^{2}$, independent of whether
the holes are localized or not.
This mechanism can naturally explain
why $D(0)$ is simply proportional to $x$, as shown in Fig.~1.
Using the Slater-Koster parameter $V_{pd\sigma}$ = 1.8 eV obtained
from the analysis of photoemission data \cite{Saitoh}, or from the TB
fit to the LDA band structure \cite{Pap}, we yield $t_{pd} =
-(\sqrt{3}/2)V_{pd\sigma}$ = $-$1.56 eV \cite{McMahan}.
Taking $\Delta_{ct}$ = 4.5 eV \cite{Saitoh,Qui,Jung}, $t_{pd}$ =$-$1.56 eV,
$C$ = 1, and $x$ =
0.3, we obtain
$D(0)$ = 0.5 eV \AA$^{2}$, which is larger than the observed
values by about three times. This is not unreasonable since the
numerical factor $C$ might be less than 1. A more quantitative
calculation of $J$ based on the $d-p$ exchange is essential to
address this issue.

It should be noted that the
ferromagnetic ``bonds'' with the exchange energy $J$ will
be randomly distributed over the
real space if the oxygen holes are localized, whereas the
distribution of the bonds will become more homogeneous when
the holes are more mobile (a motion narrowing effect). This can account
for a conventional spin-wave
dispersion in high $T_{C}$ materials \cite{Perring}
where the conductivity is high, and an unconventional magnon softening
near the zone boundary for lower $T_{C}$ compounds \cite{Baca} where
the conductivity is lower. When the oxygen holes are ordered
(charge-ordering), as observed in La$_{0.85}$Sr$_{0.15}$MnO$_{3}$
\cite{Doloc},
the ferromagnetic bonds will form a ``superlattice'', leading to a splitting
of spin-wave dispersion at a wave vector equal to the
incommensurability of the superlattice as observed \cite{Doloc}.

The above simple model would imply that the Curie
temperature $T_{C}$ should be proportional to $x$. In reality,
$T_{C}$ is strongly dependent on the
cation radius $r_{A}$ of the Re$_{1-x}$D$_{x}$ site of the perovskite
Re$_{1-x}$D$_{x}$MnO$_{3}$ even if $x$ is fixed \cite{Hwang}. Moreover, a
giant oxygen isotope shift of
$T_{C}$ has been observed in these compounds
\cite{ZhaoNature,Zhaoreview}. Therefore,
the simple $d-p$ exchange
model cannot account for these unusual phenomena. Recently,
Alexandrov and Bratkovsky \cite{Alex} have proposed that, in addition to
the $d-p$ exchange
interaction, there is a strong electron-phonon interaction that may lead
to the formation of small (bi)polarons. In the paramagnetic state,
the singlet bipolarons (spin zero) are stable and the ferromagnetic
interaction
is produced by the thermally excited polarons (spin 1/2).
Thus, the Curie temperature $T_{C}$ is selfconsistently
determined by the polaron density
at $T_{C}$. Within this scenario, $T_{C}$ can strongly depend on the
electron-phonon coupling strength and the isotope mass \cite{Alex}, in
agreement
with experiment \cite{ZhaoNature,Zhaoreview}.  At zero temperature,
all the carriers will be polarons if $J_{pd}S$ is much larger than the
bipolaron binding energy $\Delta$, where $J_{pd}$ is the exchange
energy between Mn and oxygen-hole spins. In this case, $D(0)$
$\propto$ $x$, as in La$_{1-x}$Sr$_{x}$MnO$_{3}$ system.
On the other hand, there is a mixture of polarons and
bipolarons if $J_{pd}S$ is slightly larger than $\Delta$. Then
the relation $D(0)$
$\propto$ $x$ does not hold, and $D(0)$ is proportional to the density
of polarons which is less than $x$ per cell. One should also expect
that the mixture of mobile polarons and localized bipolarons would
lead to a dynamic/static phase separation since bosonic and fermionic
carriers tend to separate in real space.

Finally we would like to address whether the band structure shown in
Fig.~3 can be consistent with an effective 1-orbital DE model with a
large $t$ $\sim$ 1.2 eV. We can show that in order to
have a $T_{C}$ of 380 K with such
a $t$, one needs an unphysically large value of $J_{H}$, i.e.,
$J_{H}$ $>$ 5 eV. Since the $d-p$ exchange model can well explain
the ferromagnetism, it would not be neccessary to have such an effective DE
model for describing the physics of manganites. We believe that
the $d-p$ exchange + (bi)polaron model \cite{Alex} is sufficient to
account for the essential physics in manganites.

{\bf Acknowlegement}: We would like to thank A. S. Alexandrov, S.
Satpathy, A. Chattopadhyay, A. J. Millis, G. Khaliullin, E.
Dagotto, K. A. M\"uller, H. Keller, T. Moritomo, T. A. Kaplan and P.
C. Dai for useful discussions. The work was supported by the Swiss National
Science
Foundation.

\bibliographystyle{prsty}

\end{document}